\documentclass[a4paper,twoside]{article}

    \usepackage{amsmath,amssymb}
    \usepackage{bm}
    \numberwithin{equation}{section}

\begin{document}

    \setlength{\baselineskip}{5mm}

\begin{center}

{\LARGE Wightman-Function Approach to the Relativistic Complex-Ghost Field Theory 
}

\vspace*{5mm}

Noboru NAKANISHI
\footnote{Professor Emeritus of Kyoto University. E-mail:
nbr-nak@trio.plala.or.jp}\\

\textit{12-20 Asahigaoka-cho, Hirakata 573-0026, Japan}\\
 
\end{center}

\vspace*{5mm}

    \setlength{\baselineskip}{5mm}

{\small The relativistic complex-ghost field theory is covariantly
formulated in terms of Wightman functions. The Fourier transform of the
2-point Wightman function of a complex-ghost pair is
explicitly calculated, and its spontaneous breakdown of
Lorentz invariance is compared with that of the corresponding Feynman
integral.}

    \setlength{\baselineskip}{6mm}

\section{Introduction}

Abe and the present author have proposed and established the general method for
solving quantum field theory, formulated as the canonical operator formalism,
in the Heisenberg picture.$^{1)}$ The set of the full-dimensional (anti)commutators
for field operators is obtained as the solution 
to the q-number Cauchy problem constructed
by field equations and equal-time (anti)commutation relations, 
and then the representation of this operator solution is constructed by
giving the set of Wightman functions. In contrast to the axiomatic quantum
field theory, however, the Wightman functions thus obtained do not, in general,
satisfy the norm-positivity condition of the state-vector space.
That is, the natural framework of the Lagrangian quantum field theory
is the indefinite-metric theory.

One of the striking properties of the indefinite-metric theory
is that the eigenvalues of a hermitian operator are not necessarily
real, that is, we generally encounter complex-energy eigenvalues for 
the Hamiltonian. This fact contradicts the usual spectral
condition postulated in the axiomatic quantum field theory.
The ``energy-positivity condition" has also been made use of the above-mentioned
method for solving quantum field theory in determining Wightman functions.
That is, we have required every Wightman function $W(x_1,\, ... \, ,x_n)$
to be a boundary value of an analytic function from the lower half-planes 
of the time differences $x_1^0 - x_2^0, \, ... \, ,x_{n-1}^0 -x_n^0$;
this property is what follows from the positivity of energy.
Hence, if there are complex-energy states, we must examine whether or not
the ``energy-positivity conditon" is still applicable to our method.  

The complex-energy states should not appear in the physical world,
that is, they should be unphysical states. Hence they are called the ``complex
ghosts".$^{2)}$ A single complex ghost has zero norm, and it cannot appear in
the final state if the initial state is a real-energy state because of 
the energy conservation. However, if a complex ghost exists, there is always a
complex-conjugate ghost, and the state consisting of a pair of a complex ghost
and its conjugate has real energy and can have negative norm.
Hence, the energy conservation cannot forbid the appearance of a
complex-ghost pair and it is expected that the unitarity of the
physical S-matrix is violated. This fact was indeed confirmed in
nonrelativistic models.

In 1969-1970, however, Lee and Wick$^{3),4)}$ (and Lee$^{5)}$ alone) pointed out that
the violation of the unitarity of the physical S-matrix does not occur
in the \textit{relativistic} complex-ghost field theory. This is because as
the expression for the relativistic energy is a square root of a
sum of a mass squared and a spatial momentum squared, the complex 
conjugation of a complex mass does not generically imply the complex conjugation 
of the corresponding energy; therefore, the unitarity cut due to the complex-ghost
pair is absent. This wonderful result, however, is achieved at the
cost of the violation of Lorentz invariance, as was proved
 by the present author$^{6)}$
and as was explicitly confirmed by Gleeson, Moore, Rechenberg, and 
Sudarshan$^{7)}$ (GMRS) subsequently.
Lee and Wick$^{8)}$ attributed the violation of Lorentz invariance to the fact that
spatial momentum is real while energy is complex, and wished to 
admit the use of complex spatial momenta by adopting the
S-matrix-theoretical rule proposed by Cutkosky, Landshoff, Olive,
and Polkinghorn$^{9)}$ previously.
The violation of Lorentz invariance, however, is \textit{not} a direct
consequence of the reality of spatial momenta, because the
complex energy is encountered only on the mass shell and quantum field
theory is formulated on the basis of the off-the-mass-shell quantities.
Indeed, the present author$^{10)}$ showed that the relativistic complex-ghost
field theory can be formulated manifestly covariantly without using
complex masses in the action and that it is actually Lorentz invariant
at the operator level.

The purpose of the present paper is to reformulate the relativistic 
complex-ghost field theory in terms of Wightman functions. Its 
manifestly covariant formulation done previously was based on the 
momentum-space consideration:$^{10)}$ After introducing creation and annihilation
operators for real-mass fields explicitly, the complex-ghost fields
were constructed by means of the Bogoliubov transformation. In the
present paper, we directly solve the Cauchy problem for the 4-dimensional
commutators and then construct the Wightman functions without going
to the momentum space. We show that the ``energy-positivity condition"
works well in spite of the presence of complex ghosts. The 2-point
Wightman functions of the fundamental fields are shown to be Lorentz invariant.
The violation of Lorentz invariance, however, takes place for the 
2-point Wightman functions of composite fields. We explicitly
calculate the Fourier transform  of the 2-point Wightman function 
of a complex-ghost pair,
and compare it with the corresponding Feynman integral calculated
by GMRS.  Although no
spectral representation holds in the present case, the expression for
the Wightman function is found to be something like the spectral function
for the Feynman amplitude.

The present paper is organized as follows. In \S2, we apply the 
Wightman-function approach to the complex-ghost theory and find
the Wightman functions explicitly. In \S3, we review the spontaneous breakdown of
Lorentz invariance in the Feynman integral involving a complex-ghost-pair
intermediate state. In \S4, we explicitly calculate the
Fourier transforms of the 2-point Wightman functions
of a complex-ghost pair and of two complex ghosts. The final section is
devoted to discussion.  

\section{Formulation of the theory}

We start with the Lagrangian density$^{5),10)}$
\begin{equation} 
\mathcal{L}=\frac{1}{2} \sum_{j=1}^2 (-1)^{j-1} (\partial^\mu
\phi_j \cdot \partial_\mu \phi_j - m^2 \phi_j^{\; 2})
-\gamma m \phi_1 \phi_2,
\end{equation}
where $\phi_1$ and $\phi_2$ are hermitian scalar fields and $m$ and $\gamma$
are positive constants.

Field equations are
\begin{equation}
 \begin{split}
(\square + m^2)\phi_1 + \gamma m \phi_2 &= 0, \\
(\square + m^2)\phi_2 - \gamma m \phi_1 &= 0,
 \end{split}
\end{equation}
and non-vanishing equal-time commutation relations are
\begin{equation}
[\partial_0 \phi_j (x), \, \phi_j (y)]_{x^0 = y^0} = -(-1)^{j-1} i\delta
(\bm{x} -\bm{y}).
\end{equation}
We set 
\begin{equation}
[\phi_j (x), \, \phi_k (y)] \equiv i\varDelta_{jk} (x-y)
\end{equation}
and 
\begin{equation}
\bm{\varDelta} \equiv \mathrm{matrix}(\varDelta_{jk}).
\end{equation}
We then have the following Cauchy problem:
\begin{equation}
(\square^x + m^2 + i\gamma m \sigma_2) \bm{\varDelta} (x-y) =0
\end{equation}
together with
\begin{equation}
 \begin{split}
\bm{\varDelta}(x-y)|_{x^0 = y^0} &= 0, \\
\partial_0^x \bm{\varDelta} (x-y)|_{x^0 = y^0} &= -\sigma_3 \delta(
\bm{x}-\bm{y}),
 \end{split}
\end{equation}
where $\sigma_i$ denotes the Pauli matrix.

We diagonalize (2.6) by using a unitary matrix $U\equiv(i+\sigma_1)/\sqrt{2}$. 
Because $U\sigma_2 U^{-1}=\sigma_3$ and 
$U\sigma_3 U^{-1}= -\sigma_2$, we obtain
\begin{equation}
(\square^x + m^2 + i\gamma m \sigma_3) \hat{\bm{\varDelta}} (x-y) =0
\end{equation}
together with
\begin{equation}
 \begin{split}
\hat{\bm{\varDelta}}(x-y)|_{x^0 = y^0} &= 0, \\
\partial_0^x \hat{\bm{\varDelta}} (x-y)|_{x^0 = y^0} &= \sigma_2 
\delta (\bm{x}-\bm{y}),
 \end{split}
\end{equation}
where $\hat{\bm{\varDelta}} \equiv U \bm{\varDelta} U^{-1}$.

We define the complex-mass $\varDelta$-function by the Cauchy problem
\begin{equation}
(\square^x + M^2) \varDelta(x-y; M^2) = 0,
\end{equation}
\begin{equation}
 \begin{split}
\varDelta(x-y; M^2)|_{x^0 = y^0} &= 0, \\
\partial_0^x \varDelta(x-y; M^2)|_{x^0 =y^0} &= - \delta(\bm{x} -\bm{y}),
 \end{split}
\end{equation}
where $M^2$ is a complex number such that $\Re M^2 >0$. 
The explicit expression for the complex-mass
$\varDelta$-function is given by$^{2)}$
\begin{equation}
 \begin{split}
\varDelta(\xi; M^2) &= \frac{1}{(2\pi)^3} \int d\bm{p} \frac{\sin(
\bm{p}\bm{\xi} - E_{\bm{p}}\xi^0)}{E_{\bm{p}}} \\
&= -\frac{1}{2\pi}\epsilon(\xi^0) \left[ \delta(\xi^2)
- \frac{M^2}{2}\theta(\xi^2) \frac{J_1(M\sqrt{\xi^2})}{M\sqrt{\xi^2}} \right],
 \end{split}
\end{equation}
where $E_{\bm{p}} \equiv \sqrt{M^2 + \bm{p}^2}$ and $J_1$ denotes 
a Bessel function. The complex-mass $\varDelta$-function is, of course, 
Lorentz invariant, though it behaves exponentially
as $\xi^2 \to \infty$.
 
The solution to the Cauchy problem (2.8) with (2.9) is given by
\begin{equation}
\hat{\bm{\varDelta}} (x-y) = i\sigma_+ \varDelta(x-y; M^2)
- i\sigma_- \varDelta(x-y; M^{*2}),
\end{equation}
where $\sigma_\pm \equiv (\sigma_1 \pm i\sigma_2)/2$ and $M^2\equiv
m^2 + i\gamma m$. Hence we have
\begin{equation}
\bm{\varDelta} (x-y) = U^{-1} \hat{\bm{\varDelta}}(x-y) U
=\sigma_3 \Re \varDelta(x-y; M^2) - \sigma_1 \Im \varDelta(x-y; M^2).
\end{equation}

We proceed to considering the representation of the above operator solution.
We introduce the vacuum $|0 \rangle$ and set $\langle 0| 0 \rangle = 1$.
The 1-point Wightman functions $\langle 0|\phi_j (x)|0 \rangle$
are set equal to zero so as to be consistent with translational invariance
and field equations. Then the 2-point truncated
Wightman functions are the same as the untruncated ones.

The 2-point Wightman functions $\langle 0|\phi_j (x) \phi_k (y)|0
\rangle$ must be constructed so as to be consistent with the commutator
functions given by (2.14) and with the energy-positivity condition.
We define the ``positive-energy" complex-mass $\varDelta$-function by
\begin{equation}
\varDelta^{(+)}(\xi; M^2) \equiv \frac{1}{(2\pi)^3} \int d\bm{p}
\frac{\exp i(\bm{p}\bm{\xi} - E_{\bm{p}}\xi^0)}{2E_{\bm{p}}}.
\end{equation}
Note that the momentum integral is convergent because $\Im E_{\bm{p}}$
tends to zero as $|\bm{p}| \to \infty$. 
Evidently, $\varDelta^{(+)}(\xi; M^2)$ is Lorentz invariant and has the properties
\begin{equation}
\varDelta^{(+)}(\xi; M^2) + \varDelta^{(+)}(-\xi; M^2)
= i\varDelta(\xi; M^2),
\end{equation}
\begin{equation}
\varDelta^{(+)}(-\xi; M^2) = [\varDelta^{(+)}(\xi; M^{*2})]^*.
\end{equation}
Furthermore, because $\Re E_{\bm{p}} > 0$, $\varDelta^{(+)}(\xi; M^2)$
is a boundary value of an analytic function from the lower
half-plane of $\xi_0$ in conformity with the requirement of the energy-positivity
condition.

From (2.14), we find that the 2-point (truncated) Wightman functions are
given by
\begin{equation}
\langle 0|\phi_j (x) \phi_j (y)|0 \rangle
= (-1)^{j-1} \frac{1}{2}[\varDelta^{(+)}(x-y; M^2) +
\varDelta^{(+)}(x-y; M^{*2})], 
\end{equation}
\begin{equation}
\langle 0|\phi_1 (x) \phi_2 (y)|0 \rangle
= \langle 0|\phi_2 (x) \phi_1 (y)|0 \rangle =
- \frac{1}{2i} [\varDelta^{(+)}(x-y; M^2)-\varDelta^{(+)}(x-y; M^{*2})].
\end{equation}
Of course, all higher-point truncated Wightman functions vanish.
Thus, all Wightman functions of the fundamental fields are Lorentz invariant.

\section{Violation of Lorentz invariance}

As we have shown in \S 2, the theory is strictly Lorentz invariant
as far as the Wightman functions of the fundamental fields are concerned. 
Quite interestingly, however, Lorentz invariance is 
no longer valid for composite fields.
In this section, we briefly review the results obtained previously for
the 1-loop Feynman integral involving a complex-ghost-pair intermediate state.

In the Fourier representation of the Feynman propagator $\varDelta_F (\xi; M^2)$
(but not $\varDelta_F (\xi; M^{*2})$), we need to introduce a \textit{complex}
contour $C$, which runs \textit{above} the pole located at $p_0 = \sqrt{M^2+
\bm{p}^2}$ in spite of the fact that it lies on the upper half-plane. 
Therefore, in order to have the momentum-space Feynman integral, we encounter 
the position-space integrals of the following type:
\begin{equation}
f(k_0) \equiv \frac{1}{2\pi} \int_{-\infty}^{\infty} dx^0 e^{\pm i(k_0
- \lambda) x^0},
\end{equation} 
where $\lambda$ is a \textit{complex} number. 
Naively, this integral is divergent and 
mathematically meaningless. By adopting a Gaussian adiabatic hypothesis
in defining the Dyson S-matrix, however, we can show that (3.1) should be defined 
in the sense of the ``complex $\delta$-function",$^{10)}$ $f(k_0) = \delta_c (k_0
-\lambda)$. \footnote{This notion was
introduced first by the present author$^{11)}$ in 1958.}@
This notion is defined by extending the definition of 
the Schwartz distribution in the 
following way: Let $\varphi (k_0)$ be a test function, which is an arbitrary
function holomorphic in an appropriate strip domain around the real axis; then  
\begin{equation}
\int_{-\infty}^{\infty} dk_0 \varphi (k_0) \delta_c (k_0 -\lambda)
\equiv \frac{-1}{2\pi i} \oint dk_0 \frac{\varphi (k_0)}{k_0-\lambda},
\end{equation}
where the contour goes around $k_0=\lambda$ 
anticlockwise. Of course, if $\lambda$ is
real, the complex $\delta$-functon reduces to the ordinary $\delta$-function.
 
Now, the Feynman integral involving a complex-ghost-pair 
intermediate state is given by
\begin{equation}
F_{MM^*} (p) \equiv \frac{1}{(2\pi)^4 i} \int d \bm{q} \int_C dq_0
\frac{1}{(q^2 - M^2)[(p-q)^2 - M^{*2}]},
\end{equation}
where we understand that its ultraviolet divergence is appropriately
subtracted. The integrand of (3.3) has four poles in the $q_0$ plane.
The contour $C$ runs from $-\infty$ to $+\infty$, passing through
below the two poles $-E_{\bm{q}}$ and $p_0 - E_{\bm{p}-\bm{q}}^*$
and above the other two poles $E_{\bm{q}}$ and $p_0 + E_{\bm{p}-\bm{q}}^*$.
Carrying out the contour integration of (3.3), we obtain
\begin{equation}
F_{MM^*}(p) = \frac{1}{2(2\pi)^3} \int d\bm{q} 
\left( \frac{1}{E_{\bm{q}}} + \frac{1}{E_{\bm{p}-\bm{q}}^*} \right) 
\frac{1}{p_0^{\, 2} - (E_{\bm{q}} + E_{\bm{p}-\bm{q}}^*)^2}.
\end{equation}

When $\bm{q}$ runs over the whole three-dimensional space, the locus of
\begin{equation}
p_0 = E_{\bm{q}} + E_{\bm{p}-\bm{q}}^*,
\end{equation}
which corresponds to the unitarity cut in the real-mass case,
sweeps a \textit{2-dimensional} fish-shaped domain $D$ on the $p_0$ plane. 
Its boundary curve $\varGamma$ intersects the real axis at only one point
$p_0 = 2 \, \Re E_{\bm{p}/2}$.
Evidently, the quantitiy $s_{\mathrm{min}}$
is not Lorentz invariant, where
\begin{equation} 
s_{\mathrm{min}} \equiv 4(\Re E_{\bm{p}/2})^2 - \bm{p}^2  \mspace{30mu}
(s_{\mathrm{min}} < (M+M^*)^2
\mspace{10mu}  \mathrm{for} \mspace{5mu} \bm{p} \neq 0).
\end{equation}

GMRS explicitly carried out the momentum integration
of (3.4). Their result is \footnote{In (3.7), an overall factor 1/2 has been 
corrected. This error was committed in (A3) of their paper.$^{7)}$}
\begin{equation}
 \begin{split}
F_{MM^*}(p) = &\frac{1}{16\pi^2} \, \Re \int_{\varGamma_s}
\frac{ds'}{s-s'} \\
&\left[ \frac{\sqrt{\bm{p}^2} (M^2 - M^{*2})}
{s'\sqrt{s'+\bm{p}^2}}
+ \frac{\sqrt{(s'-(M+M^*)^2)(s'-(M-M^*)^2)}}{s'} \right],
 \end{split}
\end{equation}
where $s \equiv p_0^{\; 2} -\bm{p}^2$ and the contour 
$\varGamma_s$ is the lower half part of the image of 
$\varGamma$ by the mapping $s'= p_0^{\; 2} - \bm{p}^2$;
it runs from $s_{\mathrm{min}}$ to $\infty$.

\section{Wightman function of a complex-ghost pair}

In this section, we explicitly calculate the Fourier transform of the
2-point Wightman function of a complex-ghost pair. Its \textit{formal}
expression is given by 
\begin{equation}
W_{MM^*}(p) \equiv \int d^4 \xi \, \varDelta^{(+)} (\xi; M^2)
\varDelta^{(+)} (\xi; M^{*2}) e^{ip\xi}.
\end{equation}
Of course,  naively (4.1) is meaningles because $\varDelta^{(+)}(\xi; M^2)$ is exponentially increasing in its absolute value.
Substituting (2.15) into (4.1) and carrying out one of spatial 
momentum integrations, we obtain
\begin{equation}
W_{MM^*}(p) = \frac{1}{(2\pi)^3} \int d\bm{q} \frac{1}{4E_{\bm{q}}
E_{\bm{p-q}}^*} \int_{-\infty}^{\infty} d\xi^0
\exp [i(p_0 - E_{\bm{q}} - E_{\bm{p-q}}^*)\xi^0].
\end{equation}

Now, the energy factor inside the exponential of (4.2)
is complex for $\bm{p} \neq 0$, so that the integration over 
$\xi^0$ is meaningless naively as stated above. We define the integral
over $\xi_0$ by means of the complex $\delta$-function as in \S3.
That is, according to (3.2), we consider
\begin{equation}
\int_{-\infty}^{\infty} dp_0 \, \varphi(p_0) W_{MM^*} (p)
= \oint dp_0 \, \varphi(p_0) I(p),
\end{equation}
where 
\begin{equation}
I(p) \equiv \frac{i}{4(2\pi)^3} \int d\bm{q} \frac{1}
{E_{\bm{q}}E_{\bm{p-q}}^* (p_0 - E_{\bm{q}} - E_{\bm{p-q}}^*)}.
\end{equation}
Without loss of generality, we can take the coordinate system defined
by $\bm{p} =(0, 0, p_3 >0)$. Employing the cylindrical coordinates for $\bm{q}$ 
and setting $\rho^2 = q_1^{\, 2} + q_2^{\,2}$, we obtain
\begin{equation}
I(p) = \frac{i}{32\pi^2} \int_{-\infty}^{\infty} dq_3
\int_0^{\infty}  \frac{d\rho^2}{E(\rho^2, q_3) E^*(\rho^2, -q_3)
[p_0 - E(\rho^2, q_3)- E^*(\rho^2, -q_3)]},
\end{equation}
where
\begin{equation}
E(\rho^2, q_3) \equiv \sqrt{M^2 + \rho^2 +\left(\frac{p_3}{2}+q_3 \right)^2}.
\end{equation}
By transforming the integration variable $\rho^2$ into a complex variable
\begin{equation}
q_0 =  E(\rho^2, q_3)+ E^*(\rho^2, -q_3),
\end{equation}
the integral is remarkably simplified into
\begin{equation}
I(p) = \frac{i}{16\pi^2} \int_{-\infty}^{\infty} dq_3
\int_{\alpha(q_3)}^{\infty} \frac{dq_0}{q_0(p_0 - q_0)},
\end{equation}
where
\begin{equation}
\alpha(q_3) \equiv  E(0, q_3)+ E^*(0, -q_3).
\end{equation}
The integration over $q_0$ is easily carried out; we obtain
\begin{equation}
I(p)= \frac{i}{16\pi^2 p_0} \int_{-\infty}^{\infty} dq_3
\log \frac{\alpha(q_3)-p_0}{\alpha(q_3)}. 
\end{equation}
Transforming the integration variable $q_3$ into $\alpha=\alpha(q_3)$
for $q_3 \geqq 0$ and $\alpha=\alpha(-q_3) = [\alpha(q_3)]^*$
for $q_3 \leqq 0$, we have
\begin{equation}
\begin{split}
I(p) &= \frac{i}{16\pi^2 p_0} \left( \int_{\varGamma^{(+)}}
+\int_{\varGamma^{(+)*}} \right) d\alpha
\frac{dq_3(\alpha)}{d\alpha}[\log(\alpha - p_0) - \log \alpha] \\
&= \frac{-i}{16\pi^2 p_0} \left( \int_{\varGamma^{(+)}}
+\int_{\varGamma^{(+)*}} \right) 
d\alpha \, q_3(\alpha) \left(
\frac{1}{\alpha-p_0} - \frac{1}{\alpha} \right) + c.
\end{split}
\end{equation}
Here, $q_3(\alpha)$ is the inverse function of $\alpha = \alpha(q_3)$;
explicitly,
\begin{equation}
q_3(\alpha) = \frac{p_3(M^2 - M^{*2}) + \alpha \sqrt{(\alpha^2-p_3^{\,2}-(M+M^*)^2)
(\alpha^2-p_3^{\,2}-(M-M^*)^2)}}{2(\alpha^2-p_3^{\,2})},
\end{equation}
where the sign of the square root has been chosen in such a way that $q_3=\alpha/2$ 
when $M=0$ and $q_3>p_3 /2$ as seen from (4.9); $c=i/16\pi^2$;
the contour $\varGamma^{(+)}$ is the image of the positive real axis by the mapping 
$\alpha=\alpha(q_3)$ for $q_3 \geqq 0$,
that is, it is the upper boundary curve of $D$ introduced in \S3. 
Both $\varGamma^{(+)}$ and $\varGamma^{(+)*}$ 
can be deformed into a real interval; that is, 
\begin{equation}
I(p) = \frac{-i}{8\pi^2 p_0} \int_{2\Re E_{\bm{p}/2}}^{\infty} 
d\alpha \, q_3(\alpha) \left(
\frac{1}{\alpha-p_0} - \frac{1}{\alpha} \right) + c.
\end{equation}

Now, we substitute (4.13) into (4.3):
\begin{equation}
\begin{split}
\int_{-\infty}^{\infty} dp_0 \, \varphi(p_0) W_{MM^*} (p)
&= \oint dp_0 \, \varphi(p_0) \left[ \frac{-i}{8\pi^2}
\int_{2\Re E_{\bm{p}/2}}^{\infty} 
d\alpha \, \frac{q_3(\alpha)}{\alpha(\alpha - p_0)} + c \, \right] \\
&= \int_{2\Re E_{\bm{p}/2}}^{\infty} d\alpha \, \varphi(\alpha)
\frac{1}{4\pi} \frac{q_3(\alpha)}{\alpha}.
\end{split}
\end{equation}
Rewriting the right-hand side of (4.14) into
\begin{equation}
\int_{-\infty}^{\infty} dp_0 \,
\varphi(p_0) \frac{1}{4\pi} \frac{q_3(p_0)}{p_0} 
\theta(p_0 -2\Re E_{\bm{p}/2}),
\end{equation}  
we find that
\begin{equation}
W_{MM^*}(p)=\frac{1}{4\pi} \frac{q_3(p_0)}{p_0} 
\theta(p_0 -2\Re E_{\bm{p}/2}).
\end{equation}
From (4.16) with (4.12), therefore, our final result is
\begin{equation} 
W_{MM^*}(p) = \frac{1}{8\pi} \left[ \frac{\sqrt{\bm{p}^2} (M^2 - M^{*2})}
{s\sqrt{s+\bm{p}^2}}
+ \frac{\sqrt{(s-(M+M^*)^2)(s-(M-M^*)^2)}}{s} \right]
\theta(s-s_{\mathrm{min}})
\end{equation}
for the general value of $\bm{p}$. 

Thus, $W_{MM^*}(p)$ is not only Lorentz non-invariant but also
complex-valued. In contrast to the real-mass case, in which 
the 2-point Wightman function is the 
discontinuity function of the Feynman amplitude according to
the Cutkosky rule, $W_{MM^*}(p)$ has no direct connection with $F_{MM^*}(p)$.
Nevertheless, comparing (4.17) with (3.7), we find that
the relationship between both expressions are quite
analogous to that in the real-mass case.  

For completeness, we calculate the Wightman function of two complex ghosts,
\begin{equation}
\begin{split}
W_{MM}(p) &\equiv \int d^4 \xi \, [\varDelta^{(+)} (\xi; M^2)]^2 e^{ip\xi} \\
&= \frac{1}{(2\pi)^3} \int d\bm{q} \frac{1}{4E_{\bm{q}}
E_{\bm{p-q}}} \int_{-\infty}^{\infty} d\xi^0
\exp [i(p_0 - E_{\bm{q}} - E_{\bm{p-q}})\xi^0].
\end{split}
\end{equation}
The calculation is carried out almost in the same way except for the fact that
the locus of $p_0 = E_{\bm{q}} + E_{\bm{p}-
\bm{q}}$ does not intersect the real axis. Therefore, the
final result cannot be expressed in terms of the ordinary function or distribution.
Introducing the notion of the ``complex $\theta$-function" by
\begin{equation}
\theta_c(k_0 - \lambda) \equiv \int_{-\infty}^{k_0} dk_0' \,
\delta_c (k_0' - \lambda),
\end{equation}
we can write the final result in the following way:
\begin{equation}
W_{MM} (p) = \frac{1}{8\pi} \sqrt{\frac{s-4M^2}{s}} \theta_c (s-4M^2).
\end{equation}

Of course, as $\Im M \to 0$, both (4.17) and (4.20)
tend to the well-known expression
for the Wightman function (which equals the discontinuity function
of the Feynman amplitude along the real axis) in the equal-real-mass case.

\section{Discussion}

In the present paper, we have successfully applied 
the method for solving quantum field theory 
in the Heisenberg picture to the covariant formulation of the 
complex-ghost theory. In spite of the fact
that there is complex energy spectrum , we see that 
the energy-positivity condition can be used without trouble.

We have discussed the spontaneous breakdown of Lorentz invariance encountered 
in the relativistic complex-ghost theory from the viewpoint of the
Wightman function. We have explicitly calculated the Fourier transform
of the 2-point Wightman function of a complex-ghost pair and  that of 
two complex ghosts. The former indeed exhibits the violation of Lorentz
invariance. In spite of the fact that the spectral representation 
does not hold for the Feynman amplitude, the relation between the 
Feynman amplitude and the corresponding Wightman function is 
quite analogous to that in the real-mass case. 

We have seen that the introduction of the complex $\delta$-function is quite essential
in those considerations. From (2.18), we have
\begin{equation}
\langle 0|[\phi_j (x)]^2 [\phi_j (y)]^2|0 \rangle
= \frac{1}{2} [\varDelta^{(+)}(x-y; M^2)+\varDelta^{(+)}(x-y; M^{*2})]^2.
\end{equation}
According to (4.17) and (4.20), therefore, its Fourier transform
is neither Lorentz invariant nor expressible in terms of the ordinary
distributions. This result is rather surprising in view of the 
fact that the Lagrangian
density (2.1) is a very simple one and explicitly contains no
complex numbers. That means that it is very difficult to avoid the appearance
of such possibilities \textit{a priori} in the generic framework of the
Lagrangian quantum field theory.
\\

{\small
\begin{center}
\textbf{References}\\
\end{center}

1) \; For a review, see N. Nakanishi, Prog.~Theor.~Phys.~\textbf{111} (2004), 301.
Further references are 
\hspace*{8mm} contained therein.

2) \; For a review, see N. Nakanishi, Prog.~Theor.~Phys.~Suppl. No.51 (1973), 1. Further 
\hspace*{8mm} references are contained therein.

3) \; T. D. Lee and G. C. Wick, Nucl.~Phys.~\textbf{B9} (1969), 209.

4) \; T. D. Lee and G. C. Wick, Phys.~Rev.~\textbf{D2} (1970), 1033.

5) \; T. D. Lee, \textit{Quanta: Essays in Theoretical Physics Dedicated to Gregor Wentzel} (Chicago 
\hspace*{8mm} U. P., Chicago, 1970), p.260.

6) \; N. Nakanishi, Phys.~Rev.~\textbf{D3} (1971), 811.

7) \; A. M. Gleeson, R. J. Moore, H. Rechenberg, and E. C. G. Sudarshan,
Phys.~Rev.~\textbf{D4} (1971), 
\hspace*{8mm} 2242.

8) \; T. D. Lee and G. C. Wick, Phys.~Rev.~\textbf{D3} (1971), 1046.

9) \; R. E. Cutkosky, P. V. Landshoff, D. I. Olive, and J. C. Polkinghorne,
Nucl.~Phys.~\textbf{B12} 
\hspace*{8mm} (1969), 281.

10) \; N. Nakanishi, Phys.~Rev.~\textbf{D5} (1972), 1968.

11) \; N. Nakanishi, Prog.~Theor.~Phys.~\textbf{19} (1958), 607.

\end{document}